\documentclass[conference]{IEEEtran}
\IEEEoverridecommandlockouts
\usepackage{cite}
\usepackage{amsmath,amssymb,amsfonts,epsfig}
\usepackage{algorithmic}
\usepackage{graphicx}
\usepackage{textcomp}
\usepackage{xcolor}
\usepackage{stackengine}
\def\BibTeX{{\rm B\kern-.05em{\sc i\kern-.025em b}\kern-.08em
    T\kern-.1667em\lower.7ex\hbox{E}\kern-.125emX}}

\newcommand{\beq}{\begin{equation}}
\newcommand{\eeq}{\end{equation}}
\newcommand{\beqn}{\begin{eqnarray}}
\newcommand{\eeqn}{\end{eqnarray}}
\DeclareMathOperator*{\argmin}{arg\,min}
\def\bmath#1{\mbox{\boldmath$#1$}}

\begin{document}

\title{Polarization-based online interference mitigation in radio interferometry\\
\thanks{This work is supported by Netherlands eScience Center (project DIRAC, grant 27016G05).}
}

\author{\IEEEauthorblockN{Sarod Yatawatta}
\IEEEauthorblockA{\textit{ASTRON, The Netherlands Institute for Radio Astronomy,} \\
Dwingeloo, The Netherlands.\\
yatawatta@astron.nl}
}

\maketitle

\begin{abstract}
Mitigation of radio frequency interference (RFI)  is essential to deliver science-ready radio interferometric data to astronomers. In this paper, using dual polarized radio interferometers, we propose to use the polarization information of  post-correlation interference signals to detect and mitigate them. We use the directional statistics of the polarized signals as the detection criteria and formulate a distributed, wideband spectrum sensing problem. Using consensus optimization, we solve this in an online manner, working with mini-batches of data. We present extensive results based on simulations to demonstrate the feasibility of our method.
\end{abstract}


\section{Introduction}
Terrestrial radio telescopes are always affected by radio frequency interference (RFI). Numerous methods have been developed for the elimination of such signals from radio interferometric data, e.g., \cite{Leshem_2000,Leshem_2000A,Fridman,raza2002spatial,Bentum2008,aoflagger,Baan2019,Cucho2019,Vos2019}. However, new sources of RFI are still emerging, e.g., \cite{Brentjens2016,Winkel2019,Soko2016}  and therefore it is important to further improve RFI mitigation techniques. Furthermore, the amount of data produced by modern radio interferometers keep increasing and therefore it is also important to develop RFI mitigation techniques that can work online, as opposed to the majority of methods that work off-line.

In this paper, we consider post-correlation RFI mitigation of radio interferometric data that are obtained by dual polarized receivers. A case in point is the low frequency array (LOFAR) \cite{LOFAR} which has dual, linearly polarized receivers. The element beam pattern of LOFAR is strongly polarized along directions close to the horizon \cite{Bregman}. Moreover, most RFI transmitters are vertically aligned on Earth \cite{Bentum2008} in stark contrast to the LOFAR receivers that lie almost flat on the ground. Therefore, RFI signals received in such a situation will have a strong polarization  signature. In spite of this, some celestial sources such as the Sun will also have strong polarization and because of this, we assume strong celestial sources are subtracted from the data before RFI mitigation is performed. Using online calibration \cite{DSW2019,Y2020}, we can subtract the signals from celestial sources in an online manner and we perform RFI mitigation as a follow up to online calibration.

Polarization state is already being used for spectrum sensing in wireless communications \cite{Guo2013,guo2016review}. In particular, we follow the method developed in \cite{Guo2013} that measures the alignment of the polarization of the RFI signal for its mitigation. In order to do this, we use directional statistics \cite{Fisher1953,Stephens1967} or statistics on the sphere. Most existing RFI mitigation techniques use the energy of the RFI signal as a detection criterion so the detection threshold directly depends on the RFI signal and noise power levels. In contrast, the proposed method uses the directionality of the RFI signal and only indirectly dependent on the RFI signal and noise power levels. Modern correlators output data covering a wide bandwidth, sampled into several thousand frequencies. In order to handle this data in an online manner, we develop a distributed, wideband spectrum sensing \cite{Quan} strategy. We also note that the signal without RFI should have a smooth and well defined behavior with frequency and the detection threshold should reflect this. Therefore, during RFI mitigation, we enforce smoothness on the detection threshold and use consensus optimization \cite{boyd2011} to find a solution.

The rest of the paper is organized as follows. We describe the signal model used for an interferometer in section \ref{sec:model}. Next, we develop a generalized likelihood ratio test (GLRT) based on directional statistics in section \ref{sec:glrt}. We provide results based on simulations in \ref{sec:simul} illustrating the performance of the proposed mitigation technique. Finally, we draw our conclusions in section \ref{sec:conclusions}.

Notation: Matrices and vectors are denoted by bold upper and lower case letters such as ${\bf J}$ and ${\bf v}$, respectively. The matrix transpose, Hermitian transpose, and pseudo-inverse are given by $(\cdot)^T$ , $(\cdot)^H$, and $(\cdot)^\dagger$ respectively. The set of real and complex numbers are denoted by  ${\mathbb R}$ and ${\mathbb C}$, respectively. The Q-function is given by $Q(\cdot)$. The matrix  Frobenius norm is given by $\|\cdot \|$.

\section{Radio interferometric data model}\label{sec:model}
The data produced by cross correlating signals from receivers $p$ and $q$ are given by \cite{HBS}
\beq
\label{V}
{\bf V}_{pq}=\sum_{i=1}^K {\bf J}_{pi} {\bf C}_{pqi} {\bf J}_{qi}^H + {\bf N}_{pq} + {\bmath \Gamma}_{pq}
\eeq
where we have $K$ signals from the sky being received. The systematic errors along direction $i$ for stations $p$ and $q$ are given by ${\bf J}_{pi}$ and ${\bf J}_{qi}$ ($\in {\mathbb C}^{2\times 2}$), respectively. The intrinsic sky signal (coherency) is ${\bf C}_{pqi}$ ($\in {\mathbb C}^{2\times 2}$). The additive, white, complex circular Gaussian noise is represented by ${\bf N}_{pq}$ ($\in {\mathbb C}^{2\times 2}$). The unwanted RFI signal is given by ${\bmath \Gamma}_{pq}$ ($\in {\mathbb C}^{2\times 2}$). Before RFI mitigation is performed, we use online calibration \cite{DSW2019} to subtract the strong signals from $K^\prime$ directions in the sky to get the residual
\beq
\label{R}
{\bf R}_{pq}={\bf V}_{pq}-\sum_{i=1}^{K^\prime} \widehat{\bf J}_{pi} {\bf C}_{pqi} \widehat{\bf J}_{qi}^H. 
\eeq
The components of ${\bf R}_{pq}$ can be represented as
\beq \label{coh}
{\bf { R}}_{pq}=\left[ \begin{array}{cc}
XX& XY\\
YX & YY 
\end{array} \right].
\eeq
Using the correlation products $XX$,$XY$,$YX$ and $YY$ ($\in {\mathbb C}$) in (\ref{coh}), we can form complex Stokes parameters as
\beqn \label{Stokes}
\mathcal{I}\buildrel\triangle\over=XX+XY,\ \mathcal{Q} \buildrel\triangle\over=XX-XY,\\\nonumber
\mathcal{U} \buildrel\triangle\over=XY+YX,\ \mathcal{V} \buildrel\triangle\over=\jmath(XY-YX).
\eeqn
From (\ref{Stokes}), we can extract either the real or the imaginary part to form conventional Stokes parameters, for instance, $I=\mathrm{real}(\mathcal{I})$, $Q=\mathrm{real}(\mathcal{Q})$, $U=\mathrm{real}(\mathcal{U})$ and $V=\mathrm{real}(\mathcal{V})$. The same can be done for the imaginary part so we can use two sets of Stokes parameters for mitigation of RFI as we explain later.

\begin{figure}[htbp]
\begin{minipage}[b]{0.98\linewidth}
\centering
\centerline{\epsfig{figure=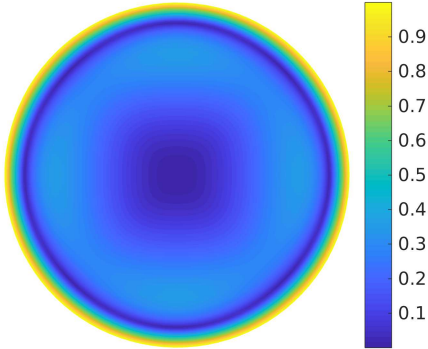,width=7.5cm}}
\end{minipage}
\caption{The LOFAR element beam polarization as a fraction of the intensity at $120$ MHz. Using the voltage beam for any given direction in the sky ${\bf E} \in\mathbb{C}^{2\times 2}$, the polarization is obtained by the components of ${\bf E}{\bf E}^H$. The full hemisphere is shown projected onto the plane. The center is pointing towards the zenith and the outer ring is the horizon.}
\label{beam}
\end{figure}

In Fig. \ref{beam}, we show the normalized polarization $\sqrt{|\mathcal{Q}|^2+|\mathcal{U}|^2+|\mathcal{V}|^2}/|\mathcal{I}|$ due to the element beam pattern of LOFAR at $120$ MHz. The increase in polarization towards the horizon is clearly seen in this figure.

Given the polarization components $(Q,U,V)$, we define the polarization vector ${\bf x}$ as
\beq \label{polvec}
{\bf x}\buildrel\triangle\over=[Q/g, U/g, V/g]^T,\ g=\sqrt{Q^2+U^2+V^2}.
\eeq
Using spherical polar coordinates, we can represent ${\bf x}$ on the Poincar\'e sphere as ${\bf x}=[\cos\theta, \sin\theta \cos\phi, \sin\theta \sin\phi]^T$, where $(\theta,\phi)$ are spherical polar coordinates.

We test two hypotheses on the distribution of ${\bf x}$, following \cite{Guo2013}. The absence or presence of RFI can be summarized as 
\beq
\mathcal{H}_0: {\bmath \Gamma}_{pq}={\bf 0}\ \ \mathrm{and}\ \  \mathcal{H}_1: {\bmath \Gamma}_{pq}\ne{\bf 0}.
\eeq

Under $\mathcal{H}_0$, we get a spherical uniform distribution
\beq \label{pdfH0}
f({\bf x} {\mid} \mathcal{H}_0)=\frac{1}{4\pi} \sin\theta
\eeq
and under $\mathcal{H}_1$, we get a Von Mises-Fisher distribution
\beq \label{pdfH1}
f({\bf x} {\mid} \mathcal{H}_1, {\bmath \mu},\kappa) =\frac{\kappa \sin\theta}{4\pi \sinh \kappa} \exp\left(\kappa {\bmath \mu}^T {\bf x}\right) 
\eeq
where ${\bmath \mu}$ is the mean direction and $\kappa$ is the concentration along that direction.

Note that while \cite{Guo2013} has derived (\ref{pdfH0}) and (\ref{pdfH1}) for auto-correlations, we re-use the same results for cross-correlations here because $p\ne q$. We consider the difference in systematics between receivers $p$ and $q$ as an effect similar to the wireless propagation model (e.g. Rayleigh fading model) used by \cite{Guo2013} to justify this re-use.

\section{Generalized likelihood ratio test}\label{sec:glrt}
Consider $N$ data points collected for baseline $pq$ in (\ref{V}), each data point being taken at a unique time and frequency. For $W$ frequencies and $T$ time samples, $N=W\times T$. Assuming independent and identically distributed data, let ${\bf X}=({\bf x}_1,\ldots,{\bf x}_N)$. The likelihood ratio between $\mathcal{H}_1$ and $\mathcal{H}_0$ is given by
\beq \label{lr}
\frac{f\left({\bf X}{\mid}\mathcal{H}_1,{\bmath \mu},\kappa \right)}{f\left({\bf X}{\mid}\mathcal{H}_0\right)} = \frac{\prod_i f({\bf x}_i {\mid} \mathcal{H}_1, {\bmath \mu},\kappa)}{\prod_i f({\bf x}_i {\mid} \mathcal{H}_0)}.
\eeq
In order to evaluate (\ref{lr}), we need to find ${\bmath \mu}$ and $\kappa$ in (\ref{pdfH1}). The maximum likelihood (ML) estimate for ${\bmath \mu}$ is given by
\beq \label{MLmu}
 {\bf R}\buildrel\triangle\over=\sum_i {\bf x}_i,\ R\buildrel\triangle\over=|{\bf R}|,\ \widehat{\bmath \mu}=\frac{\bf R}{R}
\eeq
where ${\bf R}$ is called the resultant vector and $R$ its length. The ML estimate for $\kappa$ satisfies
\beq \label{MLk}
 A(\kappa)\buildrel \triangle\over=\frac{I_{\frac{3}{2}}(\kappa)}{I_{\frac{1}{2}}(\kappa)} = \frac{ R}{N} = \overline{R}
\eeq
where $I_{j}(\cdot)$ is the modified Bessel function of the first kind and order $j$.
We do not have a closed form solution for (\ref{MLk}) but we can use a few Newton-Raphson iterations  \cite{dhillon2003} with initial value 
\beq
\kappa^0=\frac{3\overline{R}-\overline{R}^3}{1-\overline{R}^2}
\eeq
and
\beq
 \kappa^{k+1}=\kappa^{k}-\frac{A(\kappa^k)-\overline{R}}{1-A(\kappa^k)^2-\frac{2}{\kappa^k} A(\kappa^k)},
\eeq
for $k=0,1,\ldots$ to find the ML estimate of $\kappa$.

Thereafter, the likelihood ratio test can be reduced to
\beq \label{lrt}
\frac{1}{N} {\bmath \mu}^T \sum_i {\bf x}_i\ \  \underset{\mathcal{H}_0}{\overset{\mathcal{H}_1}{\gtrless}}\ \ \gamma_r
\eeq
and with the ML estimates we get,
\beq \label{glrt}
R\ \  \underset{\mathcal{H}_0}{\overset{\mathcal{H}_1}{\gtrless}}\ \  \gamma
\eeq
as the GLRT ($\gamma_r$ and $\gamma$ are pre-defined thresholds).

In order to find $\gamma$ in (\ref{glrt}), we need to measure the performance of the GLRT. We use asymptotic expressions for the probabilities using \cite{Guo2013} but exact expressions \cite{Fisher1953,Stephens1967} can be used for better accuracy.
The probability of false alarm is approximately given by
\beq \label{Pf}
P_f(\gamma)=2 Q\left(\sqrt{\frac{3}{N}\gamma}\right)+ \sqrt{\frac{6}{\pi N}\gamma}\exp\left(-\frac{3\gamma^2}{2N}\right)
\eeq
and the probability of detection is approximately given by
\beq \label{Pd}
P_d(\gamma)=Q\left(\frac{\gamma-N(\coth \kappa - 1/\kappa)-1/\kappa}{\sqrt{N(1/\kappa^2-{\mathrm{cosech}}^2 \kappa)-1/\kappa^2}}\right).
\eeq
Note that $P_d(\gamma)$ is dependent on the data (via $\kappa$) while $P_f(\gamma)$ is only dependent on $N$. Using (\ref{Pd}), the probability of missed detection is obtained as $1-P_d(\gamma)$.

We consider data at $M$ frequencies, divided into $\frac{M}{W}$ windows, and the window length in time samples is $T$. In off-line RFI mitigation, $T$ can be very large to cover the full duration of the observation and $M$ is generally smaller because the data are divided into subbands in frequency and stored at different locations. In contrast, during online (and distributed) calibration, we work with the data at all $M$ frequencies but calibration solutions are obtained for only a small value of $T$. This is to accommodate the rapid variation with time of ${\bf J}_{pi}$ and ${\bf J}_{qi}$ in (\ref{V}). Therefore, during online RFI mitigation, we consider $M$ to be very large and $T$ to be small. For the $i$-th window, $i=1\ldots \frac{M}{W}$, the detection threshold $\gamma_i$ is determined independently as in wideband spectrum sensing \cite{Quan}.

However, we do note that under $\mathcal{H}_0$, the noise spectrum varies smoothly. Therefore, we introduce the smoothness constraint $\gamma_i={\bf b}_i^T {\bf z}$ (${\bf b}_i$ and ${\bf z} \in \mathbb{R}^{F+1\times 1}$) where ${\bf b}_i$ is a polynomial basis which is evaluated at the center frequency of the $i$-th window ($f_i$) as
\beq \label{bpoly}
{\bf b}_i=\left[\left(\frac{f_i-f_0}{f_0}\right)^0, \left(\frac{f_i-f_0}{f_0}\right)^1,\ldots,\left(\frac{f_i-f_0}{f_0}\right)^{F}\right]^T
\eeq
and $f_0$ is the center frequency of all $M$ frequencies. The order of the polynomial is $F$.
The detection thresholds for all $\frac{M}{W}$ windows are determined as
\beqn \label{wb}
\gamma_1,\ldots,\gamma_{\frac{M}{W}}=\underset{\gamma_1,\ldots,\gamma_{\frac{M}{W}}} {\argmin}\ \sum_i P_f(\gamma_i)+1-P_d(\gamma_i)\\\nonumber
 \mathrm{subject\ to\ } \gamma_i={\bf b}_i^T {\bf z},\ \  i=1,\ldots,\frac{M}{W}. &&
\eeqn
We use consensus alternating direction method of multipliers (ADMM) \cite{boyd2011} to solve (\ref{wb}). The augmented Lagrangian is given by
\beqn \label{aug}
\lefteqn{L(\gamma_1,\ldots,\gamma_{\frac{M}{W}},y_1,\ldots,y_{\frac{M}{W}},{\bf z})}\\\nonumber
&=&\sum_i \left(P_f(\gamma_i)+1-P_d(\gamma_i) + y_i(\gamma_i-{\bf b}_i^T {\bf z})\right.\\\nonumber
&&\left. +\frac{\rho}{2}(\gamma_i-{\bf b}_i^T {\bf z})^2 \right)
\eeqn
where $\rho$ is the regularization parameter.
The ADMM iterations are given by
\beq \label{step1}
\gamma_i \leftarrow \underset{\gamma_i}{\argmin}\  L( \gamma_1,\ldots,\gamma_{\frac{M}{W}},y_1,\ldots,y_{\frac{M}{W}},{\bf z}),
\eeq
\beq \label{step2}
{\bf z} \leftarrow \left( \rho \sum_i {\bf b}_i {\bf b}_i^T \right)^{\dagger} \sum_i {\bf b}_i \left(y_i+\rho \gamma_i\right),
\eeq
and
\beq \label{step3}
y_i \leftarrow y_i+\rho(\gamma_i-{\bf b}_i^T {\bf z}).
\eeq
The steps (\ref{step1}) and (\ref{step3}) are performed in parallel at various distributed compute agents (that have the data for each window locally available) while (\ref{step2}) is performed at a fusion center. Solving (\ref{step1}) is performed as a bound constrained nonlinear optimization, initialized with $\bar{\gamma}$ using the approximate false error probability given by \cite{Guo2013} as
\beq \label{glow}
\bar{\gamma}=\sqrt{\frac{N}{3 c_2} Q^{-1}\left(\frac{\bar{P_f}}{c_1}\right)}
\eeq
where $c_1=1.856697$ and $c_2=0.283628$. In (\ref{glow}), $\bar{P_f}$ is the desired false error probability which is pre-defined. We also determine the bounds for (\ref{step1}) based on $\bar{\gamma}$, e.g., $[0.5\bar{\gamma},5\bar{\gamma}]$.

\section{Simulations}\label{sec:simul}
We simulate data taken over $400$ subbands, each having $64$ frequencies, thus $M=25\ 600$. The frequency range is from $110$ to $180$ MHz. The window size in time is $T=10$ (seconds) and in frequency is $W=4$, thus $N=40$. The total number of windows is therefore $M/W=6\ 400$. For baseline $pq$, we simulate (\ref{R}) as follows. First, we simulate noise ${\bf N}_{pq}$ by generating zero mean, complex circular Gaussian values for its entries. Due to the loss in sensitivity of the receiver at both the low and high ends of the band, the variance is increased towards both edges. The unsubtracted sky signal still present in ${\bf R}_{pq}$ is added as an additional zero mean, complex circular Gaussian noise with an inverse power law in frequency (thus increasing the variance at the low end of frequencies). 

\begin{figure}[htbp]
\begin{minipage}[b]{0.98\linewidth}
\centering
\centerline{\epsfig{figure=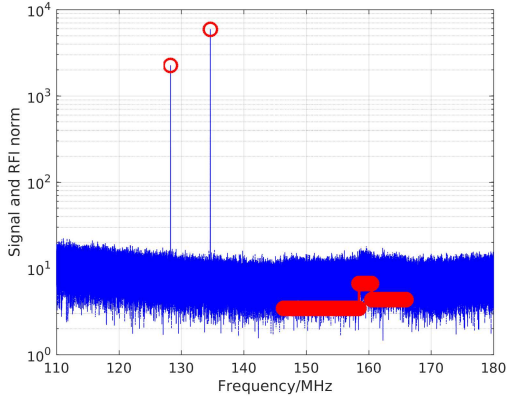,width=7.5cm}}
\end{minipage}
\caption{The signal$+$RFI norm and the RFI norm for one baseline of data. The RFI is shown in red.}
\label{vis_rfi}
\end{figure}

The RFI signal ${\bmath \Gamma}_{pq}$ is simulated by adding both narrow-band, high amplitude RFI as well as wideband, low amplitude RFI. The amplitude, the location and the width in frequency as well as the polarization of each RFI signal are randomly generated as well. The width of the narrow band RFI is kept fixed to occupy $4$ frequencies. In Fig. \ref{vis_rfi}, we show one realization of the signal $\|{\bf R}_{pq}\|$ and the RFI added to that signal $\|{\bmath \Gamma}_{pq}\|$ for all $M\times T$ data points. While we clearly see the narrow-band, high amplitude RFI, the wideband, weak RFI is hardly visible. The increase in noise variance towards the edges of the frequency range is also visible.

\begin{figure}[htbp]
\begin{minipage}[b]{0.98\linewidth}
\centering
\centerline{\epsfig{figure=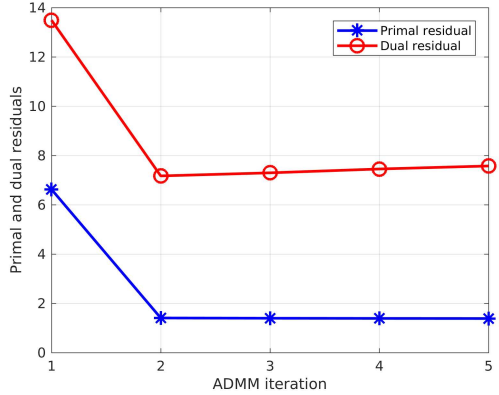,width=7.5cm}}
\end{minipage}
\caption{The variation of primal and dual residuals with ADMM iteration.}
\label{primal_dual}
\end{figure}

As we have mentioned in section \ref{sec:model}, because we have complex Stokes parameters as given by (\ref{Stokes}), we perform detections using both the real polarization components and the imaginary polarization component separately and consider either test as a positive detection in (\ref{glrt}). We perform $5$ ADMM iterations (\ref{step1}), (\ref{step2}) and (\ref{step3}) with this data. We use $F=2$ order polynomial basis with regularization $\rho=0.001$. Initial $\bar{\gamma}$ is set with $\bar{P_f}=0.01$ using (\ref{glow}). In Fig. \ref{primal_dual}, we have shown the primal residual (average of $\gamma_i - {\bf b}_i^T {\bf z}$) and dual residual (average change in ${\bf z}$) at each ADMM iteration.

\begin{figure}[htbp]
\begin{minipage}[b]{0.98\linewidth}
\centering
\centerline{\epsfig{figure=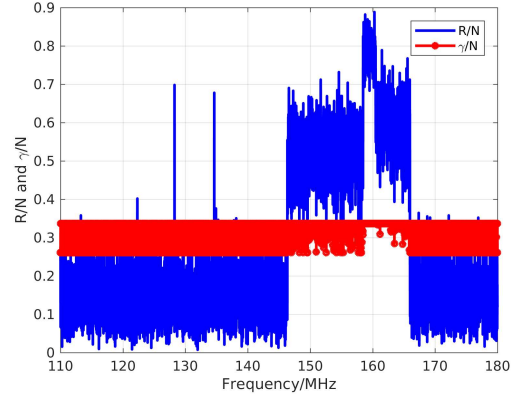,width=7.5cm}}
\end{minipage}
\caption{The GLRT quantities. Whenever $R>\gamma$, we consider $\mathcal{H}_1$ and RFI to be present.}
\label{Rgamma}
\end{figure}

In Fig. \ref{Rgamma}, we show the normalized resultant vector length $R/N$ and the normalized threshold $\gamma/N$ obtained after $5$ ADMM iterations. The data is considered RFI if $R>\gamma$ and is flagged. While the wideband RFI is not clearly visible in the signal power level in Fig. \ref{vis_rfi}, it is clearly visible (and detectable) in Fig. \ref{Rgamma}.

We perform $100$ Monte Carlo iterations with the same setup, with one exception -- i.e., we omit the simulation of narrow-band, high amplitude RFI (because it is easily detected). When simulating wideband, low amplitude RFI, we adjust the RFI power level in terms of the interference to noise ratio (INR). The INR is defined as $\frac{\|{\bmath \Gamma}_{pq}\|}{\|{\bf R}_{pq}-{\bmath \Gamma}_{pq}\|}$ and we only evaluate this using the data where RFI is present.
\begin{figure}[htbp]
\begin{minipage}[b]{0.98\linewidth}
\centering
\centerline{\epsfig{figure=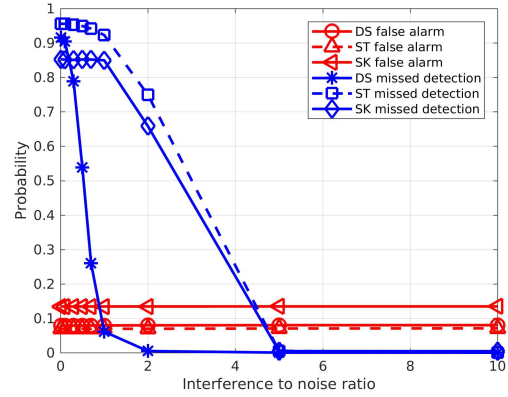,width=7.5cm}}
\end{minipage}
\caption{The variation of probability of false alarm and probability of missed detection with interference to noise ratio. DS (directional statistics): proposed method, ST: sum-threshold \cite{aoflagger}, SK: spectral kurtosis \cite{SK2010}.}
\label{SIR}
\end{figure}

In Fig. \ref{SIR}, we show the probability of false alarm $P_f(\gamma)$ as well as the probability of missed detection $1-P_d(\gamma)$  for various values of INR, averaged over $100$ simulations (for each INR). We see an almost constant false alarm probability $P_f(\gamma)$. We also see satisfactory detection of wideband, weak RFI, even at power levels close the the signal power. By increasing the window size $N$ (e.g. by combining multiple baselines), we can improve the performance even further. Furthermore, we also show the performance of two conventional RFI mitigation methods: sum-threshold \cite{aoflagger} and spectral kurtosis \cite{SK2010} in Fig. \ref{SIR}. We clearly see that the proposed method shows better performance compared with conventional methods.
\section{Conclusions}\label{sec:conclusions}
We have adopted polarization-based spectrum sensing \cite{Guo2013} in a wideband setting \cite{Quan} to develop a novel, polarization-based, online and distributed RFI mitigation algorithm for post-correlation radio interferometric data. We have shown its superior performance, even at low INR levels, using simulations. Future work will focus on the case where $\mathcal{H}_0$ will also have a Von Mises-Fisher distribution, for instance due to polarized signals from the Galaxy. 
\bibliographystyle{IEEEbib}
\bibliography{references}

\end{document}